\documentclass[twocolumn,showpacs,preprintnumbers,amsmath,amssymb]{revtex4}

\usepackage{graphicx}
\usepackage{dcolumn}
\usepackage{bm}
\usepackage{amsfonts}
\usepackage[bookmarks]{hyperref}

\newcommand{\ket}[1]{\mbox{$\mid \! #1 \, \rangle$}}
\newcommand{\bra}[1]{\mbox{$\langle \, #1 \! \mid$}}

\newcommand{\ketbra}[1]{\!\mid \! #1 \, \rangle\langle \, #1 \! \mid}

\newcommand{\eea}{\end{eqnarray}}
\newcommand{\bea}{\begin{eqnarray}}
\newcommand{\eeas}{\end{eqnarray*}}
\newcommand{\beas}{\begin{eqnarray*}}

\newcommand{\sx}{\ensuremath{\sigma_x}}
\newcommand{\sy}{\ensuremath{\sigma_y}}
\newcommand{\sz}{\ensuremath{\sigma_z}}

\newcommand{\D}{\ensuremath{D_4^{(2)}}}
\newcommand{\CC}{\ket{C}}

\begin{document}


\title{Discriminating multi-partite entangled states}

\author{Christian Schmid$^{1,2}$, Nikolai Kiesel$^{1,2}$, Wies{\l}aw Laskowski$^3$, Witlef Wieczorek$^{1,2}$, Marek \.{Z}ukowski$^3$ and Harald Weinfurter$^{1,2}$}

\affiliation{ $^{1}$Department f{\"u}r Physik,
Ludwig-Maximilians-Universit{\"a}t, D-80797 M{\"u}nchen,
Germany\\
$^{2}$Max-Planck-Institut f{\"u}r Quantenoptik, D-85748 Garching, Germany\\
$^{3}$Instytut Fizyki Teoretycznej i Astrofizyki, Uniwersytet
Gda\'{n}ski, PL-80-952 Gda\'{n}sk, Poland}%

\date{\today}

\begin{abstract}
The variety of multi-partite entangled states enables
 numerous applications in novel quantum information
tasks. In order to compare the suitability of different states
from a \emph{theoretical} point of view classifications have been
introduced. Accordingly, here we derive criteria and demonstrate
how to \emph{experimentally} discriminate an observed state
against the ones of certain other classes of multi-partite
entangled states. Our method, originating in Bell inequalities,
adds an important tool for the characterization of multi-party
entanglement.
\end{abstract}


\maketitle

Entanglement is the crucial resource for quantum information
processing and as such the "currency" to pay with in almost all
applications. For two-partite quantum states measures have been
developed that uniquely specify the value of this resource. In
contrast, for n-partite states the picture changes significantly.
First, one has to distinguish not only between fully separable or
entangled, but also between genuine n-partite, bi-, and tri-
separable entangled states, etc. Second, even states with the same
level of separability are different in the sense that they have,
for example, different Schmidt rank \cite{Ter00a} or that they
cannot be transformed into each other, e.g., by, local unitary
(LU) or, more generally, by stochastic local operations and
classical communication (SLOCC) \cite{Duer00, Ver02}. From an
experimental point of view, classifying states according to the
latter property is reasonable, as states from one SLOCC-class are
suited for the same multi-party quantum communication
applications. Thus, for the usage of multi-partite states it is of
importance to know not only the \emph{amount} but also the
\emph{type} of entanglement contained in a particular state. In
other words, the value \emph{and} the type of the "currency" is
what matters.

Tools to detect the entanglement of a state exist, most
prominently entanglement witnesses \cite{witness}. An alternative
method, relying on the correlations between results obtained by
local measurements, are Bell inequalities. 
Being originally devised to test fundamental issues of quantum
physics they allow to distinguish entangled from separable
two-qubit quantum systems \cite{Gis91,Terhal00}. Bell
inequalities, meanwhile extended to three- and more partite
quantum states \cite{Mer90,more2,ZBWW}, can thus serve as witness
for both entanglement and the violation of local realism. Recently
it was observed that for each graph state all non-vanishing
correlations (or even a restricted number thereof) form a
Bell-inequality, which is maximally violated only by the
respective quantum state \cite{Sca05, Gueh05}. In particular, the
Bell inequality for the four-qubit cluster state is not violated
at all by GHZ states \cite{Sca05}. Naturally several questions
arise: Whether one can in general apply such Bell inequalities to
discriminate particular states from other classes of multi-partite
entangled states, if so, whether they can also be constructed and
applied for non-graph states, and finally, whether there are other
operators that allow to experimentally discriminate entanglement
classes.

In this article we address these problems starting from Bell
inequalities. We present a way to construct Bell operators
\cite{Bra92} that are \emph{characteristic} for a particular
quantum state, i.e., operators that have maximal expectation value
for this
multi-partite state, only. 
With respect to experimental applications
we further aim that the expectation value can be obtained by a
minimal number of measurement settings. Under certain conditions,
we can relax the initial requirement that characteristic operators
have to be also Bell operators, which allows further reduction of
the number of settings. Comparison of the experimentally obtained
expectation values with the maximal expectation values for states
from other entanglement classes enables us to clearly distinguish
observed states from other multi-party entangled states.

In order to construct a Bell operator, we exploit the fact that
certain correlations between measurement results on individual
qubits are specific for multi-partite quantum states \cite{ZBWW}.
All correlations for a state $\ket{X}$ are summarized by the
correlation tensor $T$. If we focus on the case of four qubits,
then $T_{ijkl}=\bra{X}(\sigma_i \otimes \sigma_j \otimes \sigma_k
\otimes \sigma_l)\ket{X}$, with $i,j,k,l \in \{0,x,y,z\}$, where
$\sigma_0=\openone$ and $\sigma_{x,y,z}$ are the Pauli spin
operators. To obtain a Bell operator $\hat{\mathcal{B}}_X$ which
is characteristic for a state $\ket{X}$, we require that $\ket{X}$
is the eigenstate of $\hat{\mathcal{B}}_X$ with the highest
eigenvalue $\lambda_{\mathrm{max}}$. If the eigenstate is not
degenerate, this 
implies that $\hat{\mathcal{B}}_X$,
acting on another state cannot lead to an expectation value
greater or equal $\lambda_{\mathrm{max}}$.

An operator, which is in general not a Bell operator, but
trivially fulfills the condition to have $\ket{X}$ as the only
eigenstate with $\lambda_{\mathrm{max}}=1$, is the projector or
fidelity operator $\hat{\mathcal{F}}_X=\,\ketbra{X}$ and
\begin{equation}
\hat{\mathcal{F}}_X=\frac{1}{16} \sum_{i,j,k,l} T_{ijkl} \,
(\sigma_i \otimes \sigma_j \otimes \sigma_k \otimes \sigma_l).
\end{equation}
For most of the relevant quantum states the major part of the 256
coefficients $T_{ijkl}$ is zero. Therefore, the number of
measurement settings necessary for the evaluation of
$\hat{\mathcal{F}}_X$ is much smaller than for a complete state
tomography. We consider the non-vanishing terms as relevant
correlations for characterizing the state and take them as a
starting point for the construction of $\hat{\mathcal{B}}_X$. As
we will see in the following two examples, there are quantum
states for which a small subset of the relevant correlations is
enough to construct $\hat{\mathcal{B}}_X$. Once this is
accomplished one can calculate the upper bound, $v_Y^\ast$, on the
expectation values $v_Y=\bra{Y}\hat{\mathcal{B}}_X\ket{Y}=\langle
\hat{\mathcal{B}}_X\rangle_Y$ for states $\ket{Y}$ which belong to
other classes than $\ket{X}$. Consequently, a state under
investigation with $\langle\hat{\mathcal{B}}_X\rangle_Z=v_Z$
cannot be an element of any class of states with $v_Y^\ast<v_Z$.

Note, $\langle \hat{\mathcal{B}}_X\rangle$ induces a particular
ordering of states which is neither absolute nor related to some
entanglement of the states and, similarly to the entanglement
witness, depends on the operator $\hat{\mathcal{B}}_X$. Yet, now
we do not only detect higher or lower degree of entanglement: we
distinguish different types of entanglement. One might say that a
state with a higher $\langle \hat{\mathcal{B}}_X\rangle$ is more
"$\ket{X}$-type" entangled. The same is true for a mixed state
$\rho$ with expectation value $v_\rho =
\mathrm{Tr}[\hat{\mathcal{B}}_X \rho]=\langle \hat{\mathcal{B}}_X
\rangle_\rho$, in the sense that it cannot solely be expressed as
a mixture of pure states $\ket{Y_i}$ with $v_{Y_i}^\ast<v_\rho$,
but it has to contain contributions with a higher "X-type"
entanglement.

Summarizing, we point at the fact that one can obtain a witness of
"$\ket{X}$-type" entanglement by constructing a discrimination
operator, which has $\ket{X}$ as non-degenerate eigenvector with
the highest eigenvalue. After all, such an operator is not unique,
neither does it necessarily have to be a Bell operator. However, a
Bell operator unconditionally detects the entanglement of the
investigated state, even if the state space is not fully known.
For example, witness operators might detect a state to be
entangled though a description of measurement results based on
local realistic models, or for that purpose, based on separable
states in higher dimensional Hilbert spaces, is possible
\cite{Aci06}. If one trusts in the representation of the state, as
shown below, even more efficient operators for state
discrimination can be devised.

Let us now apply our method to the state $\ket{\Psi_{4}}$
\cite{WZ}:
\begin{eqnarray}
\ket{\Psi_{4}}&=\frac{1}{\sqrt{3}}(\ket{0011}+\ket{1100}-\frac{1}{2}(\ket{0101}\notag\\&+\ket{0110}+\ket{1001}+\ket{1010})).
\end{eqnarray}
This state was observed in multi-photon experiments \cite{Eib03}
and can be used, for example, for decoherence free quantum
communication \cite{Bou04}, quantum telecloning \cite{Mur99}, and
multi-party secret sharing \cite{Gae07}.

The fidelity operator for that state
$\hat{\mathcal{F}}_{\Psi_{4}}$ contains 40 relevant correlation
operators $(\sigma_i \otimes \sigma_j \otimes \sigma_k \otimes
\sigma_l)$, out of which 21 describe four-qubit correlations (i.e.
do not contain $\sigma_0$). Already 10 are enough to construct a
characteristic Bell operator that has $\ket{\Psi_{4}}$ as
non-degenerate eigenstate with maximum eigenvalue
$\lambda_{\mathrm{max}}=1$:
\begin{eqnarray}
6 \,\hat{\mathcal{B}}_{\Psi_{4}} &=& \sx \otimes \sy \otimes \sy \otimes \sx + \sy \otimes \sx \otimes \sy \otimes \sx \notag \\
& -& \sy \otimes \sy \otimes \sx \otimes \sx + \sx \otimes \sz \otimes \sx \otimes \sz \notag \\
&+& \sz \otimes \sx \otimes \sx \otimes \sz - \sz \otimes \sz \otimes \sx \otimes \sx \nonumber \\
                 &+& \sz \otimes \sz \otimes \sz \otimes \sz - \sy \otimes \sy \otimes \sz \otimes \sz \notag \\
                 &+& \sy \otimes \sz \otimes \sy \otimes \sz + \sz \otimes \sy \otimes \sy \otimes \sz. \label{eqn:PSI4}
\end{eqnarray}
\begin{table}
\caption{\label{tab:psiviolation} Maximal expectation values
$\langle\hat{\mathcal{B}}_{\Psi_{4}}\rangle$}
\begin{ruledtabular}
\begin{tabular}{c || l | l }
State                                           & under LU    & under SLOCC \\
\hline
\ket{\Psi_{4}}                                  & 1.000       & 1.000\\
\ket{\D}                                        & 0.926       & 0.926\\
\ket{GHZ}                                       & 0.805       & 0.805\\
\CC                                             & 0.515       & 0.764\\
\ket{W}                                         & 0.736       & 0.758\\
$\ket{\textrm{bi-sep}}$                                  & 0.722
&
0.749\\
$\ket{\textrm{sep}}$              & 0.217 &0.217
\end{tabular}
\end{ruledtabular}
\end{table}
$\hat{\mathcal{B}}_{\Psi_{4}}$ can be used to discriminate an
experimentally observed state with respect to other four-qubit
states. With the chosen normalization we obtain the limit for any
local realistic theory by replacing $\sigma_i$ by some locally
predetermined values $I_i= \pm1$, leading to the inequality
$|\langle \hat{\mathcal{B}}_{\Psi_{4}} \rangle_{\mathrm{avg}} |
\leq \frac{2}{3}$. Table \ref{tab:psiviolation} shows the bounds
on the expectation value of $\hat{\mathcal{B}}_{\Psi_{4}}$ acting
on some classes of prominent four-qubit states (including a fully
separable state $\ket{\textrm{sep}}$, any bi-separable state
$\ket{\textrm{bi-sep}}$, as well as the four-partite entangled
Dicke state $D^{(2)}_4$ \cite{DICKE}, the GHZ \cite{GHZ1}, W
\cite{Duer00} and Cluster ($C$) \cite{Rau01} state). These bounds
were obtained by numerical optimization over either LU- or
SLOCC-transformations, respectively. In particular with the bound
for an arbitrary bi-separable state $\hat{\mathcal{B}}_{\Psi_{4}}$
provides also a sufficient
condition for genuine four-partite entanglement. 
\begin{table}[t]
\caption{\label{tab:dickeviolation} Maximal expectation values
$\langle\hat{\mathcal{B}}_{\D}\rangle$}
\begin{ruledtabular}
\begin{tabular}{c || l | l }
State &  under LU  & under SLOCC  \\ \hline
\ket{\D}                            &1.000&1.000\\
\ket{\Psi_{4}}                    &0.889&0.889\\
\ket{GHZ}                           &0.833&0.833\\
\CC                                 &0.500&0.706\\
$\ket{\textrm{bi-sep}}$                      &0.667&0.667\\
\ket{W}                             &0.613&0.619\\
$\ket{\textrm{sep}}$  & 0.178  &0.178
\end{tabular}
\end{ruledtabular}
\end{table}

\begin{figure}
\includegraphics[width=8.5cm,clip]{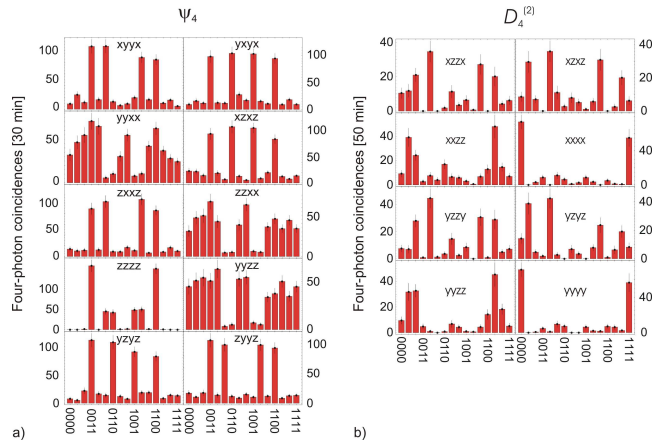}
\caption{Histogramms of the four-photon coincidence statistics for
the different measurement settings. Slots at the ordinate indicate
different events for a particular basis setting: e.g. $0011$ for
basis zzzz means detection of photons in the state $\ket{HHVV}$.
a) Statistics of the ten correlation measurements, required for
the evaluation of the operator $\hat{\mathcal{B}}_{\Psi_{4}}$. b)
Statistics of the eight correlation measurements, required for the
evaluation of the operator $\hat{\mathcal{B}}_{\D}$.}
\label{fig:w2setup}
\end{figure}

We now employ these results for the analysis of experimental data.
To observe the state $\ket{\Psi_{4}}$ we used photons generated by
type II non-collinear spontaneous parametric down conversion
(SPDC) and a variable linear optics setup. Essentially, a four
photon emission into two modes is overlapped on a polarizing beam
splitter (PBS) and subsequently split into four modes. Depending
on the setting of a half-wave plate (in our case oriented at
45$^\circ$) preceding the PBS and conditioned on detecting a
photon in each of the four outputs, a variety of states can be
observed \cite{us}. The fidelity of the experimental state
$\rho_{\Psi_{4}}$, determined from 21 four-qubit correlations, was
$\mathcal{F}_{\Psi_{4}}=\mathrm{Tr}[\hat{\mathcal{F}}_{\Psi_{4}}\rho_{\Psi_{4}}]=0.90
\pm 0.01$. The analysis of the experimental state using the Bell
operator $\hat{\mathcal{B}}_{\Psi_{4}}$ required less than half of
the measurement settings and leads to $v_{\rho_{\Psi_{4}}}=0.91
\pm 0.02$ (see Fig.~\ref{fig:w2setup}a). This value is, according
to Table \ref{tab:psiviolation}, sufficient to prove that the
experimental state is genuine four-qubit entangled and cannot be
of W-, Cluster-, or GHZ-type in the sense described above.

The class of states that can experimentally not be excluded as it has the
second largest expectation value in Table \ref{tab:psiviolation}
is represented by the so-called symmetric four qubit Dicke state
\cite{DICKE,Kie06}
\begin{eqnarray}
\ket{\D}=&\frac{1}{\sqrt{6}} (\ket{0011}+\ket{0101}+\ket{0110} \notag \\
& +\ket{1001}+\ket{1010}+\ket{1100}).
\end{eqnarray}
In turn, for the Dicke state a separate, characteristic Bell
operator $\hat{\mathcal{B}}_{\D}$ can be constructed. Again,
$\ket{\D}$ has 40 correlation operators with non zero expectation
value, out of which 21 describe original four-qubit correlations.
Naturally, the exact values of the correlations $T_{ijkl}$ differ
compared to \ket{\Psi_{4}}. In the case of $\ket{\D}$ they are
such that eight of the correlation operators are already
sufficient for the construction of $\hat{\mathcal{B}}_{\D}$:
\begin{eqnarray}\label{eqn:dop}
6 \,\hat{\mathcal{B}}_{\D} &=&  -\,\sigma_x \otimes \sigma_z
\otimes \sigma_z \otimes \sigma_x  -\sigma_x \otimes \sigma_z
\otimes \sigma_x \otimes
\sigma_z \notag \\
& & -\,\sigma_x \otimes \sigma_x \otimes \sigma_z \otimes \sigma_z
+\sigma_x \otimes \sigma_x \otimes \sigma_x \otimes \sigma_x \notag \\
& &-\,\sigma_y \otimes \sigma_z \otimes \sigma_z \otimes \sigma_y
-\sigma_y \otimes \sigma_z \otimes \sigma_y \otimes \sigma_z \notag \\
& & -\,\sigma_y \otimes \sigma_y \otimes \sigma_z \otimes \sigma_z
+\sigma_y \otimes \sigma_y \otimes \sigma_y \otimes \sigma_y ,
\end{eqnarray}
with $\lambda_{\mathrm{max}}=1$ for $\ket{\D}$. This operator has
a remarkable structure: It is of the form $\sx \otimes M_3 + \sy
\otimes M^\prime_3$, where $M_3$ and $M^\prime_3$ are three-qubit
Mermin inequality operators \cite{Mer90, footnote2}. Thus, by
applying a kind of GHZ-argument \cite{GHZ1}, the bound for any
local realistic theory can be determined to be $ |\langle
\hat{\mathcal{B}}_{\D} \rangle_{\mathrm{avg}} | \leq \frac{2}{3}$.

Table \ref{tab:dickeviolation} shows the maximal expectation
values of $\hat{\mathcal{B}}_{\D}$ by the same set of four-qubit
states as before. Considering the structure of
$\hat{\mathcal{B}}_{\D}$, further omitting correlation operators,
for example one whole block $\sx \otimes M_3$ (or $\sy \otimes
M^\prime_3$), leaves us with a four-qubit Mermin-type Bell
operator. The corresponding Bell inequality is still violated by
$\ket{\D}$. However, it is not characteristic anymore for
$\ket{\D}$ as it is maximally violated by the state
$\ket{GHZ}_y=\frac{1}{\sqrt{2}}(\ket{RRRR}\pm\ket{LLLL})$ and the
bi-separable state
$\ket{BS}=\frac{1}{\sqrt{2}}(\ket{+}(\ket{RRR}\pm i\ket{LLL}))$
(where $\ket{\pm}=\frac{1}{\sqrt{2}}(\ket{0}\pm \ket{1})$ and
$\ket{R,L}=\frac{1}{\sqrt{2}}(\ket{0}\pm i\ket{1})$ are the
eigenstates of \sx~and $\sy$, respectively). It is a particular
property of the Dicke state to have correlations in two planes
(x-z- and y-z-plane) of the Bloch sphere, whereas a GHZ state, for
instance, is correlated only in one plane (here the x-z-plane).
This quite characteristic feature is reflected in the construction
of $ \hat{\mathcal{B}}_{\D}$.
Recently, an experiment has been performed to observe the state
$\ket{\D}$ \cite{Kie06}. In order to increase the state fidelity
$\mathcal{F}$ by a higher degree of indistinguishability, here we
reduced the filter bandwidth from 3~nm to 2~nm, resulting in
$\mathcal{F}=0.92 \pm 0.02$ (compared to $\mathcal{F}=0.84 \pm
0.01$ in \cite{Kie06}). For the state's experimental analysis with
the Bell operator (\ref{eqn:dop}) we find $v_{\rho_{\D}}=0.90 \pm
0.04$ (see Fig.~\ref{fig:w2setup}b), from which we can conclude
that it is genuine four-qubit entangled and cannot be, e.g., of
W-, Cluster- or GHZ-type. Yet, this value is again just at the
limit to separate against $\ket{\Psi_{4}}$.

If one is sure about the structure of the state space, that means
that in our case it is spanned by four qubits, we can equally well
use other operators instead of the Bell operators. Let us first
drop some of the correlations from $\hat{\mathcal{B}}_{\D}$, e.g.,
the terms $(\sigma_x \otimes \sigma_x \otimes \sigma_x \otimes
\sigma_x)$ and $(\sigma_y \otimes \sigma_y \otimes \sigma_y
\otimes \sigma_y )$. The resulting discrimination operator
$\hat{\mathcal{D}}_{\D}$ is not a Bell operator anymore, but still
has $\ket{\D}$ as the only eigenstate with maximal eigenvalue
$\lambda_{\mathrm{max}}=1$ (after proper normalization).
Interestingly, as seen in Table \ref{tab:dicke6}, it introduces a
new ordering of states with a bigger separation between $\ket{\D}$
and $\ket{\Psi_{4}}$. With $v^{\mathcal{D}}_{\rho_{\D}}=0.90 \pm
0.05$ we can discriminate against this state with a better
significance. Note, the reordering, which results in the GHZ state
having now the second highest eigenvalue, indicates that this
operator analyzes the various states from a different point of
view. This is quite plausible as it uses different correlations
for the analysis. An even more radical change in the point of view
is possible with the data we dropped above, i.e., $(\sigma_x
\otimes \sigma_x \otimes \sigma_x \otimes \sigma_x)$ and
$(\sigma_y \otimes \sigma_y \otimes \sigma_y \otimes \sigma_y )$.
Relying on the particular symmetries of the Dicke state, from
these measurements we can evaluate the discrimination operator
$\hat{\mathcal{D}^\prime}_{\D}=\frac{1}{6}((\frac{1}{2}\sum_k
\sigma_x^k)^2+(\frac{1}{2}\sum_k \sigma_y^k)^2)$, where
e.g.~$\sigma_{x/y}^3=\openone \otimes \openone \otimes
\sigma_{x/y} \otimes \openone$ \cite{Tot05c}. Comparing the
observed value $v_{\rho_{\D}}^{\mathcal{D}^\prime}=0.96 \pm 0.013$
with the bounds for other states (Table \ref{tab:dicke6}) we see
that we can discriminate our state against all states of the
respective classes with only two settings. Analogous
considerations can be applied for the construction of
characteristic operators for other states \cite{Tot05b}, where the
number of settings scales polynomially with the number of qubits
compared to the exponentially increasing effort for 
state tomography.
\begin{table}
\caption{\label{tab:dicke6} Alternative characteristic operators
for $D_4^{(2)}$}
\begin{ruledtabular}
\begin{tabular}{c | c | c}
State                               &  $|\langle\hat{\mathcal{D}}_{\D}\rangle|$ (SLOCC) & $|\langle\hat{\mathcal{D}^\prime}_{\D}\rangle|$ (SLOCC) \\
\hline
\ket{\D}                            & 1.000 & 1.000 \\
\ket{GHZ}                           & 0.905 &  0.937\\
\CC                                 & 0.871 &  0.905\\
\ket{W}                             & 0.869 &  0.905\\
\ket{\Psi_{4}}                    & 0.869 &  0.901\\
$\ket{\textrm{bi-sep}}$                      &0.750 &0.872\\
$\ket{\textrm{sep}} $& 0.192 & 0.139
\end{tabular}
\end{ruledtabular}
\end{table}

In conclusion, here we showed that characteristic
\mbox{(Bell-)}operators, i.e., operators for which a particular
state only has maximal expectation value, allow to distinguish
this state from the ones out of other classes of multi-partite
entangled states. A simple, though not yet constructive, method to
design discrimination operators is based on the correlations
between local measurement settings that are typical for the
respective quantum state. The low number of measurement settings
significantly diminishes the effort compared with standard
analysis. Employing characteristic symmetries and properties of
the state under investigation can even further reduce the effort
to a number of settings which scales polynomially with the number
of qubits, thereby rendering the new method a truly efficient tool
for the characterization of multi-partite entanglement.

We thank D{.~}Bru\ss , M{.~}Horodecki, and M{.~}Wolf for
stimulating discussions. We acknowledge the support by 
the DFG-Cluster of Excellence MAP,
the DAAD/MNiSW exchange program, the 
EU Projects QAP and SECOQC. W.W. is supported by QCCC
of the ENB 
and the Studienstiftung des dt. Volkes, W.L. by FNP.


\begin{thebibliography}{32}
\expandafter\ifx\csname
natexlab\endcsname\relax\def\natexlab#1{#1}\fi
\expandafter\ifx\csname bibnamefont\endcsname\relax
  \def\bibnamefont#1{#1}\fi
\expandafter\ifx\csname bibfnamefont\endcsname\relax
  \def\bibfnamefont#1{#1}\fi
\expandafter\ifx\csname citenamefont\endcsname\relax
  \def\citenamefont#1{#1}\fi
\expandafter\ifx\csname url\endcsname\relax
  \def\url#1{\texttt{#1}}\fi
\expandafter\ifx\csname
urlprefix\endcsname\relax\def\urlprefix{URL }\fi
\providecommand{\bibinfo}[2]{#2}
\providecommand{\eprint}[2][]{\url{#2}}

\bibitem[{\citenamefont{Terhal and Horodecki}(2000)}]{Ter00a}
\bibinfo{author}{\bibfnamefont{B.~M.} \bibnamefont{Terhal}} \bibnamefont{and}
  \bibinfo{author}{\bibfnamefont{P.}~\bibnamefont{Horodecki}},
  \bibinfo{journal}{Phys. Rev. A} \textbf{\bibinfo{volume}{61}},
  \bibinfo{pages}{040301(R)} (\bibinfo{year}{2000});
\bibinfo{author}{\bibfnamefont{A.}~\bibnamefont{Sanpera}},
  \bibinfo{author}{\bibfnamefont{D.}~\bibnamefont{Bru\ss{}}}, \bibnamefont{and}
  \bibinfo{author}{\bibfnamefont{M.}~\bibnamefont{Lewenstein}},
  \bibinfo{journal}{Phys. Rev. A} \textbf{\bibinfo{volume}{63}},
  \bibinfo{pages}{050301(R)} (\bibinfo{year}{2001});
\bibinfo{author}{\bibfnamefont{Y.}~\bibnamefont{Tokunaga}},
  \bibinfo{author}{\bibfnamefont{T.}~\bibnamefont{Yamamoto}},
  \bibinfo{author}{\bibfnamefont{M.}~\bibnamefont{Koashi}}, \bibnamefont{and}
  \bibinfo{author}{\bibfnamefont{N.}~\bibnamefont{Imoto}},
  \bibinfo{journal}{Phys. Rev. A}
  \textbf{\bibinfo{volume}{74}}, \bibinfo{eid}{020301(R)}
   (\bibinfo{year}{2006}).

\bibitem[{\citenamefont{D\"ur et~al.}(2000)\citenamefont{D\"ur, Vidal, and
  Cirac}}]{Duer00}
\bibinfo{author}{\bibfnamefont{W.}~\bibnamefont{D\"ur}},
  \bibinfo{author}{\bibfnamefont{G.}~\bibnamefont{Vidal}}, \bibnamefont{and}
  \bibinfo{author}{\bibfnamefont{J.~I.} \bibnamefont{Cirac}},
  \bibinfo{journal}{Phys. Rev. A} \textbf{\bibinfo{volume}{62}},
  \bibinfo{pages}{062314} (\bibinfo{year}{2000}).

\bibitem[{\citenamefont{Verstraete et~al.}(2002)\citenamefont{Verstraete,
  Dehaene, Moor, and Verschelde}}]{Ver02}
\bibinfo{author}{\bibfnamefont{F.}~\bibnamefont{Verstraete}},
  \bibinfo{author}{\bibfnamefont{J.}~\bibnamefont{Dehaene}},
  \bibinfo{author}{\bibfnamefont{B.}~\bibnamefont{DeMoor}}, \bibnamefont{and}
  \bibinfo{author}{\bibfnamefont{H.}~\bibnamefont{Verschelde}},
  \bibinfo{journal}{Phys. Rev. A} \textbf{\bibinfo{volume}{65}},
  \bibinfo{eid}{052112} (\bibinfo{year}{2002}).

\bibitem{witness}M. Horodecki, P. Horodecki, and R. Horodecki, Phys. Lett. A {\bf 223}, 1
(1996):


\bibitem[{\citenamefont{Gisin}(1991)}]{Gis91}
\bibinfo{author}{\bibfnamefont{N.}~\bibnamefont{Gisin}},
  \bibinfo{journal}{Phys. Lett. A} \textbf{\bibinfo{volume}{154}},
  \bibinfo{pages}{201} (\bibinfo{year}{1991}).

\bibitem{Terhal00}B.M. Terhal, Phys. Lett. A \textbf{271}, 319
(2000).

\bibitem[{\citenamefont{Mermin}(1990)}]{Mer90}
\bibinfo{author}{\bibfnamefont{N.}~\bibfnamefont{D.}~\bibnamefont{Mermin}},
  \bibinfo{journal}{Phys. Rev. Lett.} \textbf{\bibinfo{volume}{65}},
  \bibinfo{pages}{1838} (\bibinfo{year}{1990}).

\bibitem[{\citenamefont{Belinski\u{i} and Klyshko}(1993)}]{more2}
\bibinfo{author}{\bibfnamefont{A.~V.} \bibnamefont{Belinski\u{i}}}
  \bibnamefont{and} \bibinfo{author}{\bibfnamefont{D.~N.}
  \bibnamefont{Klyshko}}, \bibinfo{journal}{Phys. Usp.}
  \textbf{\bibinfo{volume}{36}}, \bibinfo{pages}{653}
  (\bibinfo{year}{1993});
\bibinfo{author}{\bibfnamefont{W.}~\bibnamefont{Laskowski}},
  \bibinfo{author}{\bibfnamefont{T.}~\bibnamefont{Paterek}},
  \bibinfo{author}{\bibfnamefont{M.}~\bibnamefont{Zukowski}}, \bibnamefont{and}
  \bibinfo{author}{\bibfnamefont{C.}~\bibnamefont{Brukner}},
  \bibinfo{journal}{Phys. Rev. Lett.} \textbf{\bibinfo{volume}{93}},
  \bibinfo{eid}{200401} (\bibinfo{year}{2004});
\bibinfo{author}{\bibfnamefont{K.}~\bibnamefont{Nagata}},
  \bibinfo{author}{\bibfnamefont{W.}~\bibnamefont{Laskowski}},
  \bibinfo{author}{\bibfnamefont{M.}~\bibnamefont{Wie\'sniak}},
  \bibnamefont{and}
  \bibinfo{author}{\bibfnamefont{M.}~\bibnamefont{\.Zukowski}},
  \bibinfo{journal}{Phys. Rev. Lett.} \textbf{\bibinfo{volume}{93}},
  \bibinfo{pages}{230403} (\bibinfo{year}{2004}).

\bibitem[{\citenamefont{\.Zukowski and Brukner}(2002)}]{ZBWW}
\bibinfo{author}{\bibfnamefont{R.~F.} \bibnamefont{Werner}} \bibnamefont{and}
  \bibinfo{author}{\bibfnamefont{M.~M.} \bibnamefont{Wolf}},
  \bibinfo{journal}{Phys. Rev. A} \textbf{\bibinfo{volume}{64}},
  \bibinfo{pages}{032112} (\bibinfo{year}{2001});
\bibinfo{author}{\bibfnamefont{M.}~\bibnamefont{\.Zukowski}} \bibnamefont{and}
  \bibinfo{author}{\bibfnamefont{{\v C}.}~\bibnamefont{Brukner}},
  \bibinfo{journal}{Phys. Rev. Lett.} \textbf{\bibinfo{volume}{88}},
  \bibinfo{pages}{210401} (\bibinfo{year}{2002}).

  \bibitem[{\citenamefont{Scarani et~al.}(2005)\citenamefont{Scarani, Acin,
  Schenck, and Aspelmeyer}}]{Sca05}
\bibinfo{author}{\bibfnamefont{V.}~\bibnamefont{Scarani}},
  \bibinfo{author}{\bibfnamefont{A.}~\bibnamefont{Acin}},
  \bibinfo{author}{\bibfnamefont{E.}~\bibnamefont{Schenck}}, \bibnamefont{and}
  \bibinfo{author}{\bibfnamefont{M.}~\bibnamefont{Aspelmeyer}},
  \bibinfo{journal}{Phys. Rev. A} \textbf{\bibinfo{volume}{71}},
  \bibinfo{eid}{042325} (\bibinfo{year}{2005}).

\bibitem[{\citenamefont{G\"uhne et~al.}(2005)\citenamefont{G\"uhne, T\'oth,
  Hyllus, and Briegel}}]{Gueh05}
\bibinfo{author}{\bibfnamefont{O.}~\bibnamefont{G\"uhne}},
  \bibinfo{author}{\bibfnamefont{G.}~\bibnamefont{T\'oth}},
  \bibinfo{author}{\bibfnamefont{P.}~\bibnamefont{Hyllus}}, \bibnamefont{and}
  \bibinfo{author}{\bibfnamefont{H.~J.} \bibnamefont{Briegel}},
  \bibinfo{journal}{Phys. Rev. Lett.} \textbf{\bibinfo{volume}{95}},
  \bibinfo{pages}{120405} (\bibinfo{year}{2005});
\bibinfo{author}{\bibfnamefont{G.}~\bibnamefont{T\'oth}},
  \bibinfo{author}{\bibfnamefont{O.}~\bibnamefont{G\"uhne}}, \bibnamefont{and}
  \bibinfo{author}{\bibfnamefont{H.~J.} \bibnamefont{Briegel}},
  \bibinfo{journal}{Phys. Rev. A} \textbf{\bibinfo{volume}{73}},
  \bibinfo{pages}{022303} (\bibinfo{year}{2006}).

\bibitem[{\citenamefont{Braunstein et~al.}(2006)\citenamefont{Braunstein, Mann and Revzen}}]{Bra92}
\bibinfo{author}{\bibfnamefont{S.}~\bibfnamefont{L.}~\bibnamefont{Braunstein}},
  \bibinfo{author}{\bibfnamefont{A.}~\bibnamefont{Mann}},
 \bibnamefont{and}
  \bibinfo{author}{\bibfnamefont{M.}~\bibnamefont{Revzen}},
  \bibinfo{journal}{Phys. Rev. Lett.} \textbf{\bibinfo{volume}{68}}, \bibinfo{pages}{3259}
  (\bibinfo{year}{1992});
  \bibinfo{author}{\bibfnamefont{R.}~\bibfnamefont{F.}~\bibnamefont{Werner}},
 \bibnamefont{and}
  \bibinfo{author}{\bibfnamefont{M.}~\bibfnamefont{M.}~\bibnamefont{Wolf}},
  \bibinfo{journal}{Phys. Rev. A} \textbf{\bibinfo{volume}{61}}, \bibinfo{pages}{062102}
  (\bibinfo{year}{2000}).

\bibitem[{\citenamefont{Acin et~al.}(2006)\citenamefont{Acin, Gisin, and Masanes}}]{Aci06}
\bibinfo{author}{\bibfnamefont{A.}~\bibnamefont{Acin}},
  \bibinfo{author}{\bibfnamefont{N.}~\bibnamefont{Gisin}},
 \bibnamefont{and}
  \bibinfo{author}{\bibfnamefont{L.}~\bibnamefont{Masanes}},
  \bibinfo{journal}{Phys. Rev. Lett.} \textbf{\bibinfo{volume}{97}}, \bibinfo{pages}{120405} (\bibinfo{year}{2006}).

\bibitem[{\citenamefont{Weinfurter and \.Zukowski}(2001)}]{WZ}
\bibinfo{author}{\bibfnamefont{H.}~\bibnamefont{Weinfurter}} \bibnamefont{and}
  \bibinfo{author}{\bibfnamefont{M.}~\bibnamefont{\.Zukowski}},
  \bibinfo{journal}{Phys. Rev. A} \textbf{\bibinfo{volume}{64}},
  \bibinfo{pages}{010102(R)} (\bibinfo{year}{2001}).

\bibitem[{\citenamefont{Eibl et~al.}(2003)\citenamefont{Eibl, Gaertner,
  Bourennane, Kurtsiefer, \.Zukowski, and Weinfurter}}]{Eib03}
\bibinfo{author}{\bibfnamefont{M.}~\bibnamefont{Eibl}},
  \bibinfo{author}{\bibfnamefont{S.}~\bibnamefont{Gaertner}},
  \bibinfo{author}{\bibfnamefont{M.}~\bibnamefont{Bourennane}},
  \bibinfo{author}{\bibfnamefont{C.}~\bibnamefont{Kurtsiefer}},
  \bibinfo{author}{\bibfnamefont{M.}~\bibnamefont{\.Zukowski}},
  \bibnamefont{and}
  \bibinfo{author}{\bibfnamefont{H.}~\bibnamefont{Weinfurter}},
  \bibinfo{journal}{Phys. Rev. Lett.} \textbf{\bibinfo{volume}{90}},
  \bibinfo{pages}{200403} (\bibinfo{year}{2003});
\bibinfo{author}{\bibfnamefont{S.}~\bibnamefont{Gaertner}},
  \bibinfo{author}{\bibfnamefont{M.}~\bibnamefont{Bourennane}},
  \bibinfo{author}{\bibfnamefont{M.}~\bibnamefont{Eibl}},
  \bibinfo{author}{\bibfnamefont{C.}~\bibnamefont{Kurtsiefer}},
  \bibnamefont{and}
  \bibinfo{author}{\bibfnamefont{H.}~\bibnamefont{Weinfurter}},
  \bibinfo{journal}{Appl. Phys. B} \textbf{\bibinfo{volume}{77}},
  \bibinfo{pages}{803} (\bibinfo{year}{2003});
\bibinfo{author}{\bibfnamefont{J.-S.}~\bibnamefont{Xu}},
  \bibinfo{author}{\bibfnamefont{C.-F.}~\bibnamefont{Li}},
  \bibnamefont{and}
  \bibinfo{author}{\bibfnamefont{G.-C.}~\bibnamefont{Guo}},
  \bibinfo{journal}{Phys. Rev. A} \textbf{\bibinfo{volume}{74}},
  \bibinfo{pages}{052311} (\bibinfo{year}{2006}).

\bibitem[{\citenamefont{Bourennane et~al.}(2004)\citenamefont{Bourennane, Eibl,
  Gaertner, Kurtsiefer, Cabello, and Weinfurter}}]{Bou04}
\bibinfo{author}{\bibfnamefont{M.}~\bibnamefont{Bourennane}},
  \bibinfo{author}{\bibfnamefont{M.}~\bibnamefont{Eibl}},
  \bibinfo{author}{\bibfnamefont{S.}~\bibnamefont{Gaertner}},
  \bibinfo{author}{\bibfnamefont{C.}~\bibnamefont{Kurtsiefer}},
  \bibinfo{author}{\bibfnamefont{A.}~\bibnamefont{Cabello}}, \bibnamefont{and}
  \bibinfo{author}{\bibfnamefont{H.}~\bibnamefont{Weinfurter}},
  \bibinfo{journal}{Phys. Rev. Lett.} \textbf{\bibinfo{volume}{92}},
  \bibinfo{pages}{107901} (\bibinfo{year}{2004}).

\bibitem[{\citenamefont{Murao et~al.}(1999)\citenamefont{Murao, Jonathan,
  Plenio, and Vedral}}]{Mur99}
\bibinfo{author}{\bibfnamefont{M.}~\bibnamefont{Murao}},
  \bibinfo{author}{\bibfnamefont{D.}~\bibnamefont{Jonathan}},
  \bibinfo{author}{\bibfnamefont{M.~B.} \bibnamefont{Plenio}},
  \bibnamefont{and} \bibinfo{author}{\bibfnamefont{V.}~\bibnamefont{Vedral}},
  \bibinfo{journal}{Phys. Rev. A} \textbf{\bibinfo{volume}{59}},
  \bibinfo{pages}{156} (\bibinfo{year}{1999}).

\bibitem{Gae07}S. Gaertner, C. Kurtsiefer, M. Bourennane and H. Weinfurter, Phys. Rev. Lett. \textbf{98}, 020503
(2007).

\bibitem[{\citenamefont{Dicke}(1954)}]{DICKE}
\bibinfo{author}{\bibfnamefont{R.~H.} \bibnamefont{Dicke}},
  \bibinfo{journal}{Phys. Rev.} \textbf{\bibinfo{volume}{93}},
  \bibinfo{pages}{99} (\bibinfo{year}{1954}).

\bibitem[{\citenamefont{Greenberger et~al.}(1989)\citenamefont{Greenberger,
  Horne, and Zeilinger}}]{GHZ1}
\bibinfo{author}{\bibfnamefont{D.}~\bibnamefont{Greenberger}},
  \bibinfo{author}{\bibfnamefont{M.~A.} \bibnamefont{Horne}}, \bibnamefont{and}
  \bibinfo{author}{\bibfnamefont{A.}~\bibnamefont{Zeilinger}},
  \emph{\bibinfo{title}{Going beyond Bell's Theorem}}
  (\bibinfo{publisher}{Kluwer Academic, Dordrecht},
  \bibinfo{year}{1989});
\bibinfo{author}{\bibfnamefont{D.~M.} \bibnamefont{Greenberger}},
  \bibinfo{author}{\bibfnamefont{M.~A.} \bibnamefont{Horne}}, \bibnamefont{and}
  \bibinfo{author}{\bibfnamefont{A.}~\bibnamefont{Zeilinger}},
  \bibinfo{journal}{Am. J. Phys.} \textbf{\bibinfo{volume}{58}},
  \bibinfo{pages}{1131} (\bibinfo{year}{1990}).

\bibitem[{\citenamefont{Raussendorf and Briegel}(2001)}]{Rau01}
\bibinfo{author}{\bibfnamefont{R.}~\bibnamefont{Raussendorf}} \bibnamefont{and}
  \bibinfo{author}{\bibfnamefont{H.~J.} \bibnamefont{Briegel}},
  \bibinfo{journal}{Phys. Rev. Lett.} \textbf{\bibinfo{volume}{86}},
  \bibinfo{pages}{5188} (\bibinfo{year}{2001}).


\bibitem{us}W. Wieczorek et~al., in preparation.

\bibitem[{\citenamefont{Kiesel et~al.}(2007)\citenamefont{Kiesel, Schmid, T\'oth,
  Solano, and Weinfurter}}]{Kie06}
\bibinfo{author}{\bibfnamefont{N.}~\bibnamefont{Kiesel}},
  \bibinfo{author}{\bibfnamefont{C.}~\bibnamefont{Schmid}},
  \bibinfo{author}{\bibfnamefont{G.}~\bibnamefont{T\'oth}},
  \bibinfo{author}{\bibfnamefont{E.}~\bibnamefont{Solano}}, \bibnamefont{and}
  \bibinfo{author}{\bibfnamefont{H.}~\bibnamefont{Weinfurter}},
  \bibinfo{journal}{Phys. Rev. Lett.} \textbf{\bibinfo{volume}{98}},
  \bibinfo{pages}{063604} (\bibinfo{year}{2007}).

\bibitem{footnote2}The Bell inequality found by our method for the
symmetric six-qubit Dicke state with three excitations,
$\ket{D_6^{(3)}}$, is of the same structure: $\sx \otimes M_{5} +
\sy \otimes M^\prime_{5}$. The bound for local realistic theories
in this case is 0.4 and the expectation value for the Dicke state
is 1 compared to e.g.~0.85 for any six-qubit GHZ state.

\bibitem[{\citenamefont{T\'oth and G\"uhne}(2005{\natexlab{a}})}]{Tot05c}
\bibinfo{author}{\bibfnamefont{G.}~\bibnamefont{T\'oth}} \bibnamefont{and}
  \bibinfo{author}{\bibfnamefont{O.}~\bibnamefont{G\"uhne}},
  \bibinfo{journal}{Phys. Rev. A}
  \textbf{\bibinfo{volume}{72}}, \bibinfo{eid}{022340}
   (\bibinfo{year}{2005}{\natexlab{a}});
\bibinfo{author}{\bibfnamefont{G.}~\bibnamefont{T\'{o}th}},
  \bibinfo{journal}{J. Opt. Soc. Am. B} \textbf{\bibinfo{volume}{24}},
  \bibinfo{pages}{275} (\bibinfo{year}{2007}).

\bibitem[{\citenamefont{T\'oth and G\"uhne}(2005{\natexlab{b}})}]{Tot05b}
\bibinfo{author}{\bibfnamefont{G.}~\bibnamefont{T\'oth}} \bibnamefont{and}
  \bibinfo{author}{\bibfnamefont{O.}~\bibnamefont{G\"uhne}},
  \bibinfo{journal}{Phys. Rev. Lett.} \textbf{\bibinfo{volume}{94}},
  \bibinfo{eid}{060501}
  (\bibinfo{year}{2005}{\natexlab{b}});
\bibinfo{author}{\bibfnamefont{O.}~\bibnamefont{G\"uhne}},
  \bibinfo{author}{\bibfnamefont{C.-Y.} \bibnamefont{Lu}},
  \bibinfo{author}{\bibfnamefont{W.-B.} \bibnamefont{Gao}}, \bibnamefont{and}
  \bibinfo{author}{\bibfnamefont{J.-W.} \bibnamefont{Pan}},
  \bibinfo{journal}{Phys. Rev. A}
  \textbf{\bibinfo{volume}{76}}, \bibinfo{eid}{030305(R)}
   (\bibinfo{year}{2007}).
\end{thebibliography}

\copyright $\,$  2008 The American Physical Society
\end{document}